\documentclass[aps,pra,reprint,onecolumn,11pt,tightenlines,superscriptaddress,nofootinbib]{revtex4-2}
\usepackage[T1]{fontenc}
\usepackage{lmodern}
\usepackage{amsmath,amssymb,bm}
\usepackage{graphicx,booktabs,hyperref}
\usepackage{siunitx}
\DeclareSIUnit\gauss{G}
\DeclareSIUnit\debye{D}
\newcommand{\ket}[1]{\lvert #1\rangle}
\newcommand{\bra}[1]{\langle #1\rvert}
\newcommand{\Tr}{\operatorname{Tr}}

\begin{document}
\title{From spectroscopy to quantum dynamics: assessing reduced models of a NaCs molecular gate}
\author{Marko Gacesa}
\affiliation{Space Sciences Laboratory, University of California, 7 Gauss Way, Berkeley, California 94720, USA}
\date{\today}
\begin{abstract}
We investigate how spectroscopic calibration and basis selection affect predictions for a dipolar-exchange gate between two optically trapped $^{23}$Na$^{133}$Cs molecules. We construct a 128-state molecular Hamiltonian restricted to the $N=0,1$ rotational manifold, fit two effective optical parameters to three bright microwave transitions, and compare the resulting spectrum with four additional lines. In the coherent pair model, changing the optical polarization angle by $+0.1^\circ$ reduces the calculated gate fidelity from 94.64\% to 1.09\% at fixed carrier frequencies. Retuning the carriers and restoring the target Rabi frequencies recovers 89.31\%. The remaining sensitivity reflects changes in the multilevel pair dynamics that are not removed by target-frequency and target-Rabi recalibration. Enlarging the molecular basis produces nonmonotonic changes in fidelity, although the 11-state and 25-state calculations give closely agreeing logical maps after phase alignment. Two controls with isolated transfer probabilities near 98.45\% yield complete-gate fidelities differing by more than 11 percentage points. We relate these comparisons to a sufficient bound on the error of a projected propagator. The numerical tests quantify agreement among selected bases for a specified coherent Hamiltonian.
\end{abstract}
\maketitle

\section{Introduction}
Ultracold polar molecules combine long-lived internal states with electric dipole interactions that can couple spatially separated particles. Rotational and hyperfine degrees of freedom provide distinct possibilities for encoding quantum information, while optical and microwave fields provide access to selected transitions. These features motivated early proposals for molecular quantum computation~\cite{DeMille2002} and now support a broader program of quantum simulation and quantum information processing~\cite{Cornish2024}. Realizing that program requires control over both the preparation of individual molecules and the coherent evolution of interacting pairs. The same internal structure that provides useful storage and interaction states also creates additional pathways that a predictive dynamical model must resolve.

An essential foundation was the production of ultracold molecules in well-defined internal states. Ni \emph{et al.} demonstrated coherent transfer of weakly bound KRb molecules to their rovibrational ground state by stimulated Raman adiabatic passage~\cite{Ni2008}. Complementary theoretical work explored how magnetic Feshbach resonances can enhance optical access to deeply bound molecular levels. In particular, Gacesa \emph{et al.} studied short-pulse, two-color photoassociation of $^{6}$Li$^{87}$Rb near a Feshbach resonance, using resonance-induced singlet--triplet mixing to enhance the initial optical excitation before a second pulse transfers population to the ground electronic state~\cite{Gacesa2013}. This provides a useful connection to the present problem: external fields affect not only transition energies but also the composition of molecular states and the couplings available for coherent transfer. The present calculations concern microwave manipulation of already prepared NaCs molecules; they do not model the association process.

Coherent control within the ground electronic manifold supplies the next ingredient. Ospelkaus \emph{et al.} demonstrated preparation and coherent manipulation of molecular hyperfine states through two-photon microwave transitions involving rotationally excited states~\cite{Ospelkaus2010}. Such schemes exploit hyperfine mixing to connect otherwise weakly coupled storage states to electric-dipole-active rotational states. The separation of storage and interaction functions also appeared in the atom--molecule hybrid proposal of Kuznetsova \emph{et al.}, in which atomic qubits are temporarily converted into polar molecules to acquire a two-qubit phase through dipole interactions~\cite{Kuznetsova2010}. Although that architecture differs from a gate between pre-existing molecular qubits, it highlights a shared requirement: coherent access to the interacting states and retrieval from them are part of the quantum operation, together with the interaction interval itself.

Experiments have progressively established the interaction and trapping capabilities needed for this approach. Dipolar spin exchange was observed in lattice-confined polar molecules~\cite{Yan2013}, and subsequent experiments demonstrated controlled pairwise entanglement in optical tweezer arrays of CaF~\cite{Bao2023,Holland2023}. More recently, molecular iSWAP operations in NaCs~\cite{Picard2025} and long-lived entanglement of RbCs in magic-wavelength tweezers~\cite{Ruttley2025} have extended this control to assembled bialkali molecules. Optical trapping is itself an active component of coherent control: the rotational-state-dependent light shift can cause differential dephasing, while appropriately chosen polarization or wavelength can improve coherence~\cite{Ruttley2025,Park2023}. Consequently, a gate Hamiltonian must account for the optical environment as well as the applied microwave fields and intermolecular interaction.

For NaCs, Ni, Rosenband, and Grimes proposed a hyperfine-storage exchange-gate sequence and analyzed off-resonant molecular leakage using an explicit internal-state Hamiltonian~\cite{Ni2018}. Picard \emph{et al.} subsequently realized the molecular iSWAP gate and characterized microwave transfer errors, spectator dressing, and motional decoherence in optical tweezers~\cite{Picard2025}. These studies make clear why a nominal two-level transition need not be dynamically isolated. The microwave field used to access an interaction state also drives other rotational--hyperfine transitions, sometimes with substantially larger dipole matrix elements. Changes in magnetic field or optical polarization modify both the detunings and the complex transition amplitudes. Reproducing a few measured line centers therefore constrains only part of the information needed to predict a complete gate.

Several control strategies address related limitations. Hughes \emph{et al.} designed a shaped microwave entangling pulse for CaF and examined an analogous RbCs implementation, including sensitivity to confinement and thermal excitation~\cite{Hughes2020}. Smooth-pulse optimal control and robustness criteria have been developed for exchange-coupled molecular qubits~\cite{Bergonzoni2025}, and alternative molecular entangling protocols aim to reduce sensitivity to interaction fluctuations~\cite{LuShi2026}. More generally, suppressing leakage through pulse shaping is an established multilevel-control strategy~\cite{Motzoi2009}. These advances motivate a complementary assessment of the Hamiltonian on which a control calculation rests. A pulse optimized in a restricted space can perform well for its chosen objective while missing phase shifts, additional spectators, or errors incurred during other stages of the operation.

Here we ask how the reliability of a spectroscopically informed reduced Hamiltonian should be assessed when it predicts a complete quantum operation. NaCs provides a useful case: the single-molecule $N=0,1$ Hamiltonian is tractable, whereas propagation of a driven pair becomes increasingly expensive as additional internal states are retained. We examine the effects of target-transition recalibration, spectator selection, and control-objective choice in the experimental operating regime. We compare phase-sensitive logical maps across selected bases and relate these numerical checks to a sufficient error criterion for a specified Hamiltonian. The calculations distinguish agreement with static spectroscopy, stability under basis enlargement, and fidelity of the complete coherent operation. They establish the scope of predictions within the assumed internal-state model, while leaving motional and dissipative effects to a more complete description.

\section{Methodology}
\subsection{Validation criteria}
\label{sec:hierarchy}
\begin{figure}[t]
\centering\includegraphics[width=.93\linewidth]{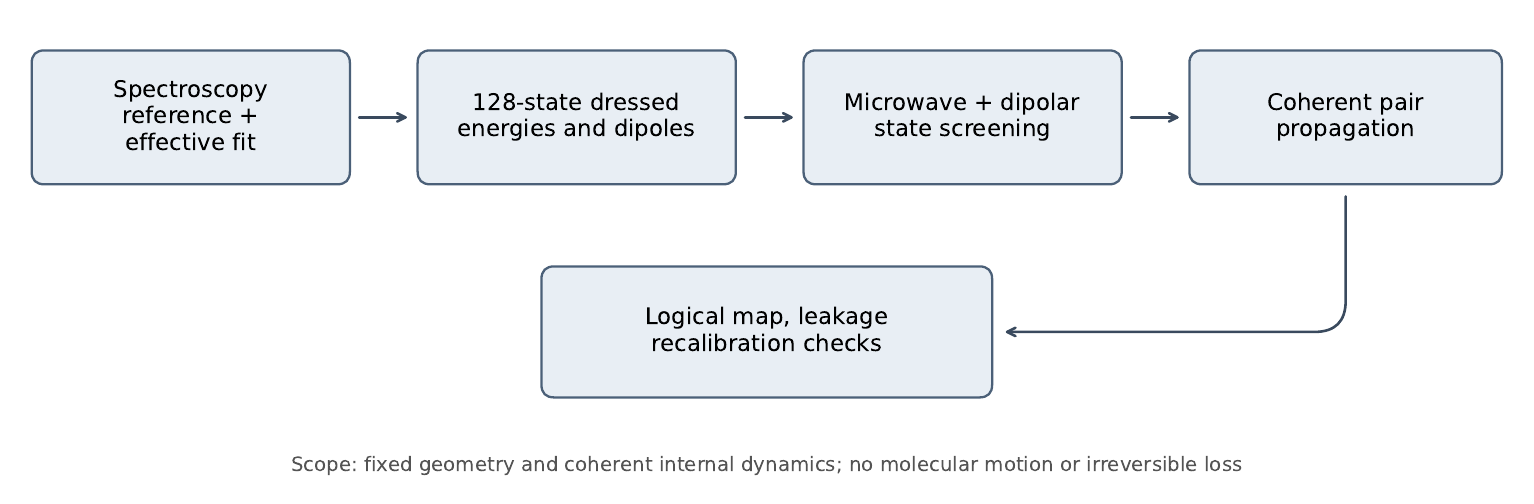}
\caption{Scope of the computational model and validation chain. Molecular reference spectroscopy constrains effective parameters of a 128-state dressed Hamiltonian. Microwave and dipolar screening define selected pair spaces for explicit coherent propagation. Comparisons are made at the level of logical maps and experimentally meaningful recalibration conventions. Motion and irreversible loss are outside the model.}
\label{fig:chain}
\end{figure}

A useful reduced model can be internally consistent and still be experimentally underdetermined. We distinguish four checks, illustrated in Fig.~\ref{fig:chain}.
\begin{enumerate}
\item \emph{Spectroscopic constraint:} Does the model reproduce transitions not used to fit its effective parameters, and are measured line centers interpreted consistently with finite-power microwave dressing?
\item \emph{Dynamical recalibration:} If a model parameter changes, which measured controls are permitted to be recalibrated? Do the target transition and all relevant spectators respond in the same way?
\item \emph{State-space stability:} Are projected amplitudes stable when omitted levels are added using more than one physical selection criterion? Does a scalar fidelity conceal differences in the complex logical map?
\item \emph{Observable and control specification:} Is the computed metric the actual requested operation, with leakage and phases included, and are its control resources and phase ramps defined explicitly?
\end{enumerate}
Each step is a necessary diagnostic for the corresponding class of error, not a certificate that a physical gate is accurately predicted. In particular, even complete convergence of an assumed coherent internal-state model would not recover unmodeled motional, stochastic, or hardware effects.

\subsection{Projected quantum dynamics}
\label{sec:projection}
We begin with a construction that applies beyond polar molecules---for example, to
truncated vibrational manifolds, multilevel atomic qubits, or coupled-spin models.
Consider a finite-dimensional Hilbert space divided into the states explicitly
retained by a simulation, with orthogonal projector $\Pi$, and all omitted states,
with $Q=\mathbb I-\Pi$. The distinction is between \emph{where the wavefunction
resides} and \emph{which states can affect its evolution}. Even when little
population remains in $Q$ at the end of an operation, virtual excursions through
$Q$ can modify the phase accumulated in $\Pi$.

To see the mechanism without reference to NaCs, write the possibly time-dependent,
Hermitian Hamiltonian as
\begin{equation}
H(t)=\begin{pmatrix} A(t)&B(t)\\B^\dagger(t)&C(t)\end{pmatrix},\quad
A=\Pi H\Pi,\quad B=\Pi HQ,\quad C=QHQ,
\label{eq:block_hamiltonian}
\end{equation}
where blocks act between the respective subspaces. Here $B$ contains the
\emph{off-block couplings}, which need not be determined by measuring a few
transition frequencies. Writing $\ket{\psi}=\ket{p}+\ket{q}$, with
$\ket{p}=\Pi\ket{\psi}$ and $\ket{q}=Q\ket{\psi}$, gives
\begin{align}
i\hbar\ket{\dot p}&=A\ket p+B\ket q,\\
i\hbar\ket{\dot q}&=B^\dagger\ket p+C\ket q.
\label{eq:coupled_pq}
\end{align}
Suppose that $\ket{q(0)}=0$. The second equation can then be solved formally
and substituted into the first. This \emph{exact} elimination yields the
projection-operator equation~\cite{Feshbach1958}
\begin{align}
i\hbar\ket{\dot p(t)}={}&A(t)\ket{p(t)}
-\frac{i}{\hbar}\int_0^t ds\, B(t)U_Q(t,s)B^\dagger(s)\ket{p(s)},
\label{eq:memory}\\
U_Q(t,s)={}&\mathcal T\exp\!\left[-\frac{i}{\hbar}
\int_s^t C(\tau)\,d\tau\right].
\end{align}
The second term describes a history-dependent process: amplitude leaves the
retained space, evolves among omitted levels, and returns. Ordinary truncation
propagates only $A(t)$ and neglects this term. The equation is valid for any
fixed choice of $\Pi$, without a weak-coupling or large-detuning assumption.

A minimal example makes the physical consequence transparent. For one retained
level and one detuned spectator, take
\begin{equation}
\frac{H}{\hbar}=\begin{pmatrix}0&\Omega/2\\\Omega/2&\Delta\end{pmatrix},
\qquad |\Omega|\ll|\Delta|,
\label{eq:two_level_example}
\end{equation}
where $\Omega$ and $\Delta$ are angular frequencies. The maximum spectator
population, starting in the retained level, is of order $(\Omega/\Delta)^2$;
yet the retained state acquires an energy shift approximately
$-\hbar\Omega^2/(4\Delta)$ and an accumulated phase of magnitude
$|\Omega^2 T/(4\Delta)|$. Thus small population transfer does not ensure
small \emph{phase} error for a long operation. This is the simplest instance
of the distinction between spectral agreement, leakage and complete coherent
dynamics; multilevel pathways add interference and time dependence.

\subsection{Residual bound for the reduced propagator}
\label{sec:criterion}
A useful criterion should refer to the \emph{dynamical observable and allowed
calibration}, rather than to basis size alone. Let
$L:\mathbb C^d\rightarrow\operatorname{Ran}(\Pi)$ be an isometric embedding
of the $d$ relevant initial logical states, $L^\dagger L=\mathbb I_d$;
$L$ need not span all of $\Pi$. Let $U(t,0)$ denote propagation under the
specified complete Hamiltonian $H(t)$, and $U_\Pi(t,0)$ propagation under
$A(t)=\Pi H(t)\Pi$, with the same control waveform. Introduce the
\emph{omitted-space residual} along the reduced trajectory,
\begin{equation}
R_L(t)=QH(t)\Pi\, U_\Pi(t,0)L,
\label{eq:residual}
\end{equation}
and its dimensionless time integral
\begin{equation}
\boxed{\ \eta_L(T)=\frac{1}{\hbar}\int_0^T
\big\|R_L(t)\big\|_2\,dt\ },
\label{eq:eta}
\end{equation}
where $\|\cdot\|_2$ is the operator norm (largest singular value).
$R_L$ is not the final leaked amplitude: it measures the instantaneous
forcing toward omitted states that the reduced calculation discards.
It can in principle be evaluated by applying the off-block Hamiltonian
to the already-computed reduced trajectories, without propagating every
omitted-state amplitude.

Duhamel's identity for unitary evolution gives the \emph{rigorous sufficient
bound}, for a fixed projector and a fully specified finite-dimensional
Hermitian $H(t)$,
\begin{align}
\big\|U(T,0)L-U_\Pi(T,0)L\big\|_2&\le\eta_L(T),
\label{eq:state_error_bound}\\
\underbrace{\frac{\big\|L^\dagger[U(T,0)-U_\Pi(T,0)]L\big\|_F}
{\sqrt d}}_{\displaystyle \delta_{\rm log}(T)}&\le\eta_L(T).
\label{eq:map_bound}
\end{align}
The Frobenius-normalized quantity $\delta_{\rm log}$ directly measures
error in the \emph{complex logical map}, with phases included in a common
reference frame. It differs from final spectator population. For the
unconditional average overlap fidelity of Eq.~\eqref{eq:metric}, both logical
maps are contractions, and the trace inequality gives
\begin{equation}
\boxed{\ \big|F_{\rm avg}^{\rm full}-F_{\rm avg}^{\Pi}\big|
\le 2\delta_{\rm log}(T)\le 2\eta_L(T)\ }.
\label{eq:fidelity_bound}
\end{equation}
This bound also applies when each model maximizes over the \emph{same set}
of permitted output phase corrections, because the maximum of functions
obeying a uniform error bound obeys that bound as well. The norm comparison
in Eq.~\eqref{eq:map_bound} must nevertheless use a consistent physical
phase frame; independently aligning maps is a diagnostic, not automatically
a gate-error certificate.

Equations~\eqref{eq:eta}--\eqref{eq:fidelity_bound} turn ``trust the model''
into a statement with a numerical tolerance. For example, a sufficient
condition for a reduction-induced change below $\epsilon_F$ in average
fidelity is $\eta_L(T)<\epsilon_F/2$. If a protocol is assessed over a
parameter neighborhood $\Theta$ and includes a specified recalibration rule
$c(\theta)$, the sufficient condition becomes
\begin{equation}
\sup_{\theta\in\Theta}\eta_L\big(T;\theta,c(\theta)\big)
<\epsilon_F/2.
\label{eq:uniform_criterion}
\end{equation}
The rule $c(\theta)$ may track a target carrier, restore a target Rabi
frequency, or preserve a fixed laboratory waveform; these are different
operational statements. This is a sufficient, \emph{not necessary}, test:
the norm integral ignores destructive interference and off-resonant
cancellation, so it may greatly overestimate the actual error. A large
$\eta_L$ does not prove that the reduction is inaccurate.

For a less conservative but noncertifying check, compare the maps from a
candidate space $\Pi$ and a systematically enlarged reference space $\Pi'$:
\begin{equation}
\delta_{\rm nest}(\theta)=\frac1{\sqrt d}
\big\|M_{\Pi'}(\theta,c(\theta))
-M_\Pi(\theta,c(\theta))\big\|_F.
\label{eq:nested_criterion}
\end{equation}
Small $\delta_{\rm nest}$ across a tested region demonstrates agreement
\emph{between those two approximations}; it is not an upper bound on their
common error relative to the complete Hilbert space. If the relevant
phase corrections are independently readjusted, one may report an
\emph{aligned} $\delta_{\rm nest}$ as a separate operational diagnostic,
with the permitted frame transformations made explicit.

\label{sec:model}
\subsection{Molecular Hamiltonian and spectroscopic calibration}
The internal basis for a single $X^1\Sigma^+$ molecule is $\ket{N,m_N,m_{\rm Na},m_{\rm Cs}}$ with $N=0,1$, $I_{\rm Na}=3/2$, and $I_{\rm Cs}=7/2$. It contains 32 ground-rotational and 96 excited-rotational basis vectors. We construct
\begin{equation}
H_{\rm mol}=H_{\rm rot}+H_{\rm quadrupole}+H_{\rm spin\text{-}rot}+H_{\rm spin\text{-}spin}+H_{\rm Zeeman}+H_{\rm opt}.
\label{eq:mol}
\end{equation}
The microscopic hyperfine terms follow the published NaCs molecular Hamiltonian~\cite{Ni2018,Aldegunde2017}, with nuclear Zeeman terms and an \emph{effective} polarization-dependent tensor light shift. At the reference magnetic field $B=\SI{864.1}{\gauss}$, two optical parameters were fitted: $u_t/h=-\SI{3526.155}{\kilo\hertz}$ and $\chi=54.817^\circ$, with the optical polarization convention $\bm\epsilon_{\rm opt}=(\cos\chi,i\sin\chi,0)$. The fit minimizes an \emph{unweighted} least-squares residual for three bright rotational transitions, keeping the theoretical hyperfine constants and magnetic field fixed (Appendix~\ref{app:fit}); it is implemented in \texttt{nacs.spectroscopy.calibrate}. The effective parameters are not a unique measurement of the tweezer field. Seven reported lines were compared; agreement of relative splittings among the three fit-constraining lines is necessarily in-sample, while the remaining static residuals rule out a claim of independently validated spectroscopy. Hypothesized finite-power microwave shifts may change observed line maxima, but the individual unknown spectroscopy powers were inferred from the discrepancies and cannot establish predictive agreement.

\begin{figure}[t]
\centering\includegraphics[width=.98\linewidth]{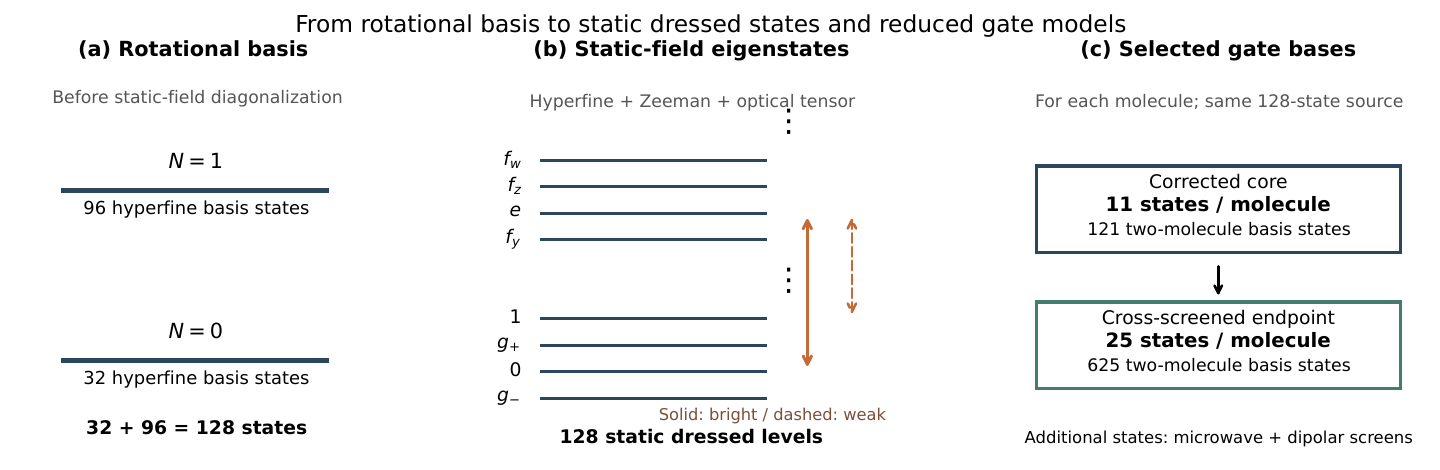}
\caption{Conceptual state-space schematic (energy spacing not to scale). (a) Rotational basis: 32 states for $N=0$ and 96 for $N=1$. (b) Representative eigenstates after diagonalizing the static hyperfine, magnetic and effective optical interactions in the full 128-state space; arrows denote possible microwave-induced couplings, not computed level splittings or microwave-power-dressed eigenstates. (c) Corrected 11-state single-molecule gate basis and selected enlargements, reaching a cross-screened 25-state endpoint (625 pair basis states). Both one-body state selection and two-body propagation are model dependent.}\label{fig:levelschematic}
\end{figure}

Figure~\ref{fig:levelschematic} distinguishes the rotational basis from the eigenstates of the \emph{static} hyperfine--Zeeman--optical Hamiltonian and indicates the reduced spaces subsequently used for gate propagation. Here ``dressed'' denotes diagonalization in the static magnetic and tweezer fields; it does not mean microwave-power-dressed quasienergies. Level positions are schematic, not calculated energies.

The dressed eigenstates $\ket{\widetilde\alpha}$ and energies $E_\alpha$ are obtained from the 128-state one-body Hamiltonian within the chosen $N=0,1$ manifold. Microwave and dipole--dipole matrix elements are constructed using the complex dressed dipole tensors
\begin{equation}
D_{q,\alpha\beta}=\bra{\widetilde\alpha}d_{\rm mol}C^{(1)}_q(\hat{\bm n})\ket{\widetilde\beta},\qquad q=-1,0,+1,
\end{equation}
with $d_{\rm mol}=\SI{4.6}{\debye}$. A single microwave antenna polarization, represented by a fixed nominal orientation of $35^\circ$, is assumed throughout the operation. That orientation is inferred in our model, not independently characterized for the apparatus. Consequently, the following sensitivity calculations are conditional rather than error bars on measured experimental parameters.

\subsection{Two-molecule dynamics and the logical observable}
For a selected one-body dressed-state basis, the pair Hamiltonian is
\begin{align}
H_{AB}(t)&=h_A(t)\otimes\mathbb I+\mathbb I\otimes h_B(t)+V_{\rm dd},\\
V_{\rm dd}&=\frac{\bm d_A\cdot\bm d_B-3(\bm d_A\cdot\hat{\bm R})(\bm d_B\cdot\hat{\bm R})}{4\pi\epsilon_0R^3}.
\label{eq:pair}
\end{align}
The geometry has a fixed $R=\SI{1.9}{\micro\meter}$ separation; excitation-nonconserving fast terms are dropped in the secular approximation. The same microwave field acts on both molecules. The modeled sequence consists of a $\ket1\leftrightarrow\ket e$ access pulse, dipolar evolution, a rotational echo, further dipolar evolution, and reverse retrieval. The interaction is retained during the driven intervals, rather than being artificially switched off during access. The numerical echo is a truncated Gaussian with parameters chosen within the model; its fitted duration should not be equated with the experimental echo duration.

Let $P_{\rm L}$ project onto the four hyperfine-storage logical states. For the full coherent operation $U(T)$, define the projected matrix $M=P_{\rm L}U(T)P_{\rm L}$, represented in the ordered logical basis. $M$ need not be unitary because population can remain outside the logical space. With $d=4$, the unconditional average overlap fidelity used in the computations is
\begin{equation}
F_{\rm avg}(Z)=\frac{\Tr(M^\dagger M)+\left|\Tr(U_{\rm target}^\dagger ZM)\right|^2}{d(d+1)},\qquad
\mathcal{L}=1-\frac{1}{d}\Tr(M^\dagger M),
\label{eq:metric}
\end{equation}
where $\mathcal{L}$ is mean final logical leakage and $Z$ is a product of allowed local output-$Z$ corrections. For fixed-parameter model comparisons, the corrections may be separately calibrated to the new static parameter set; for held-out fluctuations, nominal phase corrections are kept fixed. This distinction is essential because fitting phases independently to every fluctuation would overstate operational robustness. These metrics are neither a postselected truth-table fidelity nor an experimental process tomography estimate.

\subsection{Software and reproducibility}
\label{sec:software}
The public implementation of the molecular Hamiltonian and coherent pair propagator includes dressed energies, transition dipoles, and stage-resolved leakage diagnostics (\url{https://github.com/mgacesa66/nacs-gate-control}, version 0.1.0).

\section{Results}
\subsection{Fitted and held-out spectroscopic transitions}
Table~\ref{tab:sevenlines} lists the seven published transition offsets and the \emph{static} predictions exported by the spectroscopy baseline.

\begin{table*}[t]
\caption{Seven microwave-line offsets relative to the stated $2B_{\rm rot}$ reference, in kHz. The first three transitions were used for the unweighted two-parameter optical fit. The remaining four are held-out diagnostics. $\Delta_{\rm raw}=\nu_{\rm static}-\nu_{\rm obs}$ and $\Delta_{\rm ref}=\Delta_{\rm raw}-5.88$ kHz; the latter is a post hoc common-reference comparison, \emph{not} a Hamiltonian correction or microwave-power shift. Relative Rabi amplitudes are experimental, polarization-dependent reference quantities, not inputs to this optical fit. The published $1\to e$ Rabi entry has an unresolved normalization in the extraction and is therefore omitted.}\label{tab:sevenlines}
\begin{ruledtabular}
\begin{tabular}{llrrrrr}
Transition & Role & Measured & Static model & $\Delta_{\rm raw}$ & $\Delta_{\rm ref}$ & Rel. Rabi \\
\hline
$0\to e$ & fit & 0.200 & 6.065 & +5.865 & $-0.015$ & 1.000 \\
$0\to f_z$ & fit & 691.700 & 697.586 & +5.886 & +0.006 & 0.77 \\
$0\to f_y$ & fit & $-710.300$ & $-704.417$ & +5.883 & +0.003 & 0.84 \\
$0\to f_w$ & hold-out & 474.600 & 487.347 & +12.747 & +6.867 & 0.0184 \\
$g_+\to e$ & hold-out & 466.100 & 483.028 & +16.928 & +11.048 & 0.0053 \\
$g_-\to e$ & hold-out & $-927.100$ & $-948.012$ & $-20.912$ & $-26.792$ & 0.0218 \\
$1\to e$ & hold-out & $-1902.200$ & $-1920.389$ & $-18.189$ & $-24.069$ & --- \\
\end{tabular}
\end{ruledtabular}
\end{table*}

The reported line-center uncertainties are $0.1$--$0.3$ kHz but were not used as fit weights. The first three lines determine the effective optical parameters; the other four were withheld. After removing a common $+5.88$ kHz offset for comparison only, the in-sample bright-line residuals are $0.02$ kHz or less. The withheld weak rotational line, however, misses by $6.87$ kHz, and the hyperfine-changing lines by approximately $11$--$27$ kHz. These are discrepancies of the \emph{undriven} spectrum, not independently calculated finite-power line centers.

\subsection{Gate response to optical ellipticity and microwave recalibration}
\label{sec:calibration}
We perturb the static molecular Hamiltonian while retaining the same pulse programs, then compare three explicitly defined calibration conventions: (i) unchanged carrier frequencies and microwave electric fields, (ii) carriers retuned to the dressed target transitions, and (iii) those retunings together with electric-field rescaling that restores the target Rabi frequencies. The scan includes hypothetical ellipticity changes of $\pm0.1^\circ$, $\pm0.25^\circ$, and $\pm0.5^\circ$; additional checks use $B\pm\SI{0.1}{\gauss}$ and $\pm0.5\%$ effective tensor strength. These ranges are sensitivity probes and are not reported laboratory uncertainties.

\begin{figure}[t]
\centering\includegraphics[width=.76\linewidth]{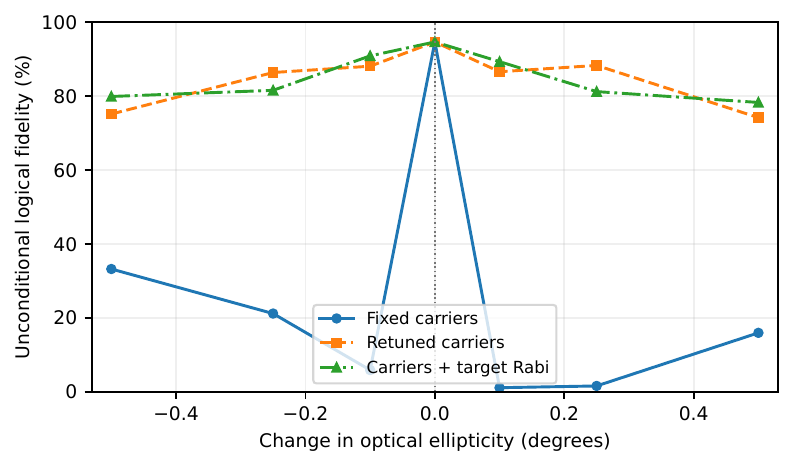}
\caption{Unconditional complete-gate fidelity under hypothetical changes of fitted optical ellipticity, with per-parameter-set static logical-$Z$ calibration. Curves share the same pulse programs. The three modes hold microwave controls fixed, track the two target carrier frequencies, or also restore target Rabi rates. Parameter intervals are not measured uncertainty bounds; improvements from retuning are idealized calibration scenarios, not experimentally demonstrated hardware responses.}
\label{fig:calibration}
\end{figure}

Figure~\ref{fig:calibration} shows the pronounced dependence of the same guarded control on this calibration convention. A $+0.1^\circ$ ellipticity change yields approximately $1.09\%$ calculated fidelity with fixed carrier frequencies, versus $86.58\%$ with target-frequency retuning and $89.31\%$ with both frequency and target-Rabi recalibration; the nominal reference is $94.64\%$. For $-0.1^\circ$, the corresponding values are $5.86\%$, $88.08\%$, and $90.87\%$. For a known static parameter change, much of the loss of fidelity with unchanged frequencies is therefore removable by recalibration. This comparison does not imply insensitivity to unknown fluctuations occurring faster than the calibration cycle.

But recalibration does not make the altered Hamiltonian equivalent to the original one. For $\pm0.5^\circ$, the corrected frequency-and-Rabi calculation remains near $78$--$80\%$ rather than the nominal $94.64\%$. The relative energy between the major spectator labeled \texttt{aux51} and the desired excited state changes by order $\SI{35}{\kilo\hertz}$ for these illustrative probes. Target-frequency and target-Rabi calibration does not, in general, restore spectator detunings and couplings or the intended dipolar-exchange matrix elements. The dwell time is held fixed in these comparisons. The observed spectator shift is evidence of a changed multilevel Hamiltonian, but does not by itself isolate the contributions of spectator excitation, exchange timing, and phase accumulation to the remaining infidelity. The response can also be nonmonotonic: independently restoring the target Rabi frequency does not improve every intermediate ellipticity point. These results show that preserving selected target-transition observables does not preserve the complete operation; they do not establish non-uniqueness of Hamiltonians fitted to the same spectroscopic data.

Static logical-phase calibration constitutes a separate layer of the same problem. For the nominal guarded protocol under $\pm\SI{0.1}{\gauss}$, target-carrier tracking produces about $94\%$ fidelity if static output $Z$ phases are refitted, but only about $38\%$ under the nominal phase convention. Most of this discrepancy is removable by static local-$Z$ recalibration, rather than by reducing leakage. Unlike a global phase, a relative local phase changes the implemented gate unless it is compensated or consistently tracked in subsequent controls. Without that correction, it remains an operational gate error. Fidelity comparisons must specify which calibrations are permitted and when they are applied.

\subsection{Selected-basis convergence and logical-map stability}
\label{sec:convergence}
A naive reduced model includes $\ket0$, $\ket1$, $\ket e$, and the most visibly coupled nearby levels. A previously used nine-state basis omitted a strong dressed spectator, labeled \texttt{aux51}, with a substantially larger microwave dipole than the weak intended hyperfine-changing transition, despite considerable detuning. That failure motivated a corrected 11-state base model and systematic screening of omitted levels from all 128 molecular eigenstates.

We use two independent screening rules. The microwave screen ranks omitted levels using drive amplitude divided by detuning across access and echo frequencies. The pair-dipolar screen ranks omitted channels by the relevant exchange matrix elements divided by two-body energy mismatch. The union of their leading candidates yields 25 states per molecule, or a 625-state pair space. The ranking is a practical truncation heuristic, not a rigorous norm bound on all omitted multistep pathways.

\begin{figure}[t]
\centering\includegraphics[width=.74\linewidth]{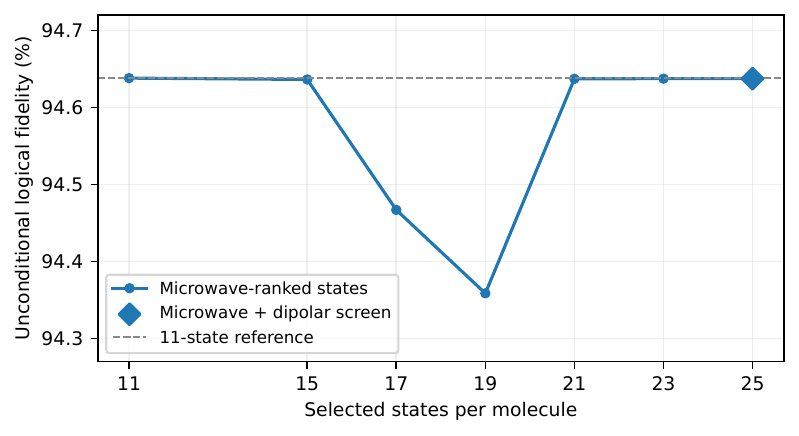}
\caption{Selected-basis convergence of the guarded pulse, with unchanged controls. The intermediate microwave-ranked 17- and 19-state calculations deviate before later states restore near-agreement with the 11-state result. A combined microwave- and dipolar-ranked 25-state endpoint provides an independent selection check. This is a tested sequence, not monotone convergence or proof of the full 128-state-per-molecule limit.}
\label{fig:basis}
\end{figure}

For a guarded control, Fig.~\ref{fig:basis} shows a nonmonotonic convergence sequence. The nominal 11-state fidelity is $94.6380\%$; the 19-state microwave-ranked space gives $94.3585\%$; and the independent-screen union at 25 states yields $94.6372\%$. Agreement of selected endpoints would conceal the intervening excursion, demonstrating why a single pairwise basis comparison is inadequate. Nevertheless, the final expanded calculations support the stability of the tested nominal observables within this family of selected spaces.

Matrix comparisons must distinguish an unobservable global phase from relative logical phases. For the nominal guarded protocol, the unnormalized Frobenius difference between the 11- and 25-state logical matrices is $0.038383$. Removing only a global phase reduces it to $0.000751$; additionally applying the allowed independently calibrated local-$Z$ corrections gives $0.000749$. For the $+0.5^\circ$ probe, the corresponding values are $0.044897$, $0.003163$, and $0.003161$. Thus the large raw differences are dominated by a global phase, and the small remaining discrepancies do not depend appreciably on local-$Z$ recalibration. These distances compare the stored logical maps, not leakage amplitudes into every omitted state.

The reported Frobenius norms are unnormalized. The aligned counterparts of Eq.~\eqref{eq:nested_criterion}, which divides by $\sqrt d=2$, are approximately $3.75\times10^{-4}$ and $1.58\times10^{-3}$, respectively. Near the perturbed operating point, two eigenstates mix strongly and both are retained; retaining their joint subspace avoids sensitivity to the assignment of individual labels near the avoided crossing.

The important claim is therefore constrained: the tested logical-map predictions survive two physical spectator screens and a substantial basis enlargement. We have not proved convergence over the entire $128^2=16{,}384$ pair basis within the chosen $N=0,1$ manifold, and spectroscopy alone does not supply the missing truncation-error bound.

\subsection{Isolated transfer and complete-gate fidelity}
\label{sec:objective}
The role of optimization is diagnostic rather than prescriptive. We compare four finite-phase-ramp control objectives with identical microwave-field and pulse-duration constraints: isolated forward/reverse transfer probabilities; complete-gate infidelity penalized by one selected spectator; complete logical-channel infidelity; and complete-channel infidelity with an all-spectator penalty. The echo and a fixed hypothetical antenna orientation are shared. The searches use common initial candidates and fixed corrected-basis refinement budgets, although one intermediate local-search stage terminates after unequal actual numbers of evaluations. None of the resulting pulses is a certified global optimum or a verified hardware design.

\begin{figure}[t]
\centering\includegraphics[width=.79\linewidth]{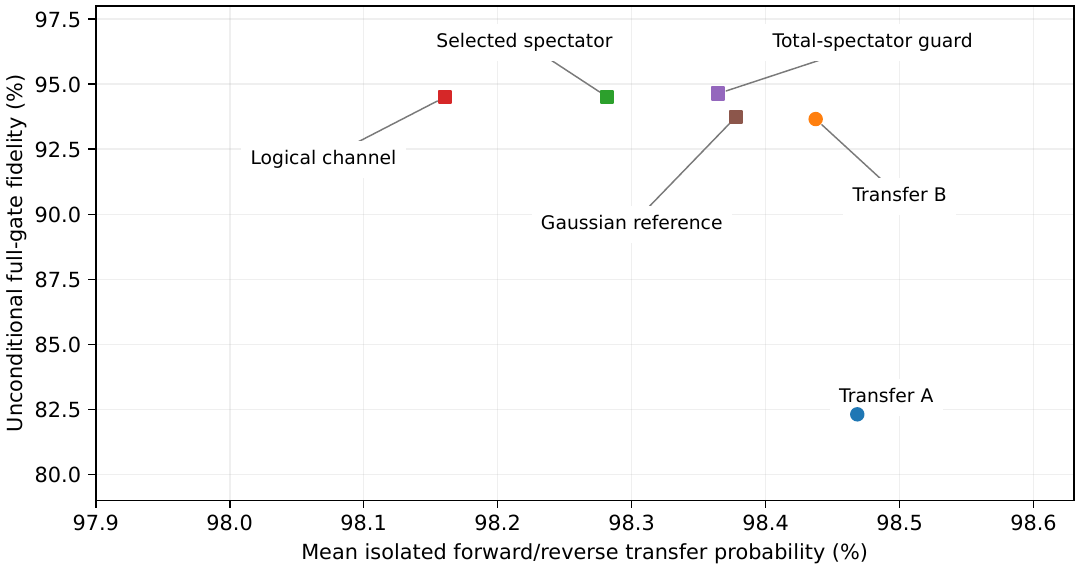}
\caption{Mean isolated forward/reverse transfer probability versus unconditional complete-gate fidelity for six stored control candidates. The two isolated-transfer optimization seeds illustrate that nearly equal transfer probabilities can correspond to very different complete logical operations because the isolated objective does not constrain all gate degrees of freedom. The remaining points show three complete-channel objectives and a Gaussian reference pulse. The plot compares model outputs, not measured gate results.}
\label{fig:objectives}
\end{figure}

As Fig.~\ref{fig:objectives} illustrates, a candidate with about $98.47\%$ isolated forward/reverse transfer probability gives only $82.31\%$ complete logical fidelity; another seed with about $98.44\%$ isolated transfer reaches $93.66\%$ full-gate fidelity. This is not a newly discovered dynamical instability: the isolated-transfer objective leaves the exchange dwell and some logical phases underconstrained. Rather, it demonstrates that a spectroscopic or single-transition benchmark does not determine the requested interacting observable.

For a single NaCs rotational qubit, Chen \emph{et al.} derived two-pulse amplitude and phase conditions for a target rotation in a rotational model that neglects hyperfine structure, and calculated leakage into higher rotational levels~\cite{Chen2026}. Those leakage channels differ from the rotational--hyperfine spectators considered here; the relevant comparison is the need to constrain both amplitudes and phases. In Fig.~\ref{fig:objectives}, the isolated-transfer objective retains only forward and reverse population probabilities. It therefore leaves both transfer phases and the subsequent exchange evolution underconstrained.

With imposed $\SI{2}{\micro\second}$ smooth phase changes, the three complete-channel objectives reach similar nominal fidelity, $94.51$--$94.64\%$ on the 11-state model, with comparable 25-state results. The all-spectator-guarded solution does not show a dramatic robustness improvement. In an effective exchange-qubit model, Bergonzoni \emph{et al.} identify restrictions on robustness with global controls and show how local controls can remove them~\cite{Bergonzoni2025}. This is relevant context for our shared-field control family, but their result does not automatically extend to this multilevel Hamiltonian, and our nominal optimizations do not test that robustness theorem. An earlier instantaneous-phase control suppressed population in \texttt{aux51} by about $92\%$ while redistributing population into other spectators and reducing complete-gate fidelity, but that large cancellation does not persist under the smoother matched-control family. We regard this contrast as a useful warning about idealized waveform families, not a basis for claiming a novel optimal-control method.

The model's response to hypothetical control errors also limits interpretation. Under tight illustrative quasi-static amplitude/detuning/exchange perturbations the complete-channel candidates have mean fidelities near $92\%$; under much broader illustrative errors the means are only about $35\%$. The assumptions are not fitted laboratory noise statistics, so these samples should not be used as performance forecasts or as evidence of a robust quantum gate.

\subsection{Leakage contribution to nominal infidelity}
The metric in Eq.~\eqref{eq:metric} permits a nonnegative decomposition of the nominal infidelity. For any allowed unitary correction $Z$, Cauchy--Schwarz gives $|\Tr(U_{\rm target}^\dagger ZM)|^2\le d\Tr(M^\dagger M)$, and hence $F_{\rm avg}\le1-\mathcal L$. Therefore
\begin{equation}
1-F_{\rm avg}=\mathcal L+\mathcal D,\qquad
\mathcal D=(1-\mathcal L)-F_{\rm avg}\ge0.
\label{eq:leakage_decomposition}
\end{equation}
Here $\mathcal D$ is the remaining logical-overlap deficit. It includes logical mismatch and state-dependent attenuation and is not a measure of pure phase error. The decomposition also holds after maximizing over the permitted local-$Z$ corrections.

\begin{table}[t]
\caption{Nominal infidelity decomposition for the three finite-phase-ramp complete-channel candidates in the 11-state model. Entries are percentages, calculated from the stored results with 64 echo steps and nominal local-$Z$ calibration. These candidates are the matched-objective calculations; the guarded reference used in the parameter and basis scans is a separately stored control.}\label{tab:leakage}
\begin{ruledtabular}
\begin{tabular}{lrrr}
Objective & $100(1-F_{\rm avg})$ & $100\mathcal L$ & $100\mathcal D$ \\
\hline
Selected-spectator penalty & 5.478 & 4.730 & 0.748 \\
Complete-channel fidelity & 5.495 & 5.109 & 0.386 \\
Guarded complete-channel & 5.356 & 4.616 & 0.740 \\
\end{tabular}
\end{ruledtabular}
\end{table}

Final logical leakage accounts for approximately $86$--$93\%$ of the nominal infidelity of these candidates (Table~\ref{tab:leakage}). This identifies the dominant contribution to this metric for the stored controls, but does not establish a fundamental fidelity ceiling or assign the leakage to a particular pulse stage. Because the evolution is coherent, final leakage is also distinct from irreversible loss and need not equal the maximum population outside the logical space during the sequence.

\subsection{Interpretation of the residual criterion for NaCs}
The residual criterion puts our numerical checks on a common footing.
First, changing the optical parameters changes both $A(t)$ and the
omitted-space forcing $R_L(t)$; retuning a target microwave transition
changes the forcing again. Therefore, the calibration scenarios
cannot be assessed solely from their new line centers. Second, the
nonmonotonic basis sequence is a practical test of
Eq.~\eqref{eq:nested_criterion}. The nominal guarded protocol has an
11-to-25-state aligned normalized distance of approximately
$3.75\times10^{-4}$; for the illustrative $+0.5^\circ$
ellipticity case it is approximately $1.58\times10^{-3}$.
Those are useful small \emph{pairwise} discrepancies, but their alignment
and the absence of a complete-space bound prevent interpretation as
certificates on the entire $128^2$-dimensional pair evolution within the chosen rotational manifold. Third,
the objective comparison illustrates why $L$ and the chosen gate observable must be specified:
a bound or convergence test for one isolated transfer need not control
the complete two-molecule sequence.

For NaCs we report pairwise differences between selected bases. The residual bound applies within a specified Hamiltonian. Uncertainty in the microwave field, trapping motion, and loss require separate physical estimates.

\section{Conclusions}
\label{sec:discussion}
The case study shows that several different forms of agreement must not be conflated. A Hamiltonian can reproduce fitted relative spectral splittings yet have uncertain microwave matrix elements. For known static changes, a large gate sensitivity may be removable by retuning carriers, Rabi rates, and local phases. Such calibration does not generally restore the surrounding spectrum or the exchange dynamics, and it does not remove unknown fast fluctuations. A selected basis may agree with a smaller basis only after intermediate nonmonotonic corrections. In the stored basis comparisons, the large raw logical-matrix differences are almost entirely a global phase; after its removal, close agreement remains without appreciable reliance on local-$Z$ adjustments.

A useful claim of \emph{model reliability} should therefore be attached to a precisely defined observable, operating neighborhood, calibration protocol, basis-selection rule, and numerical tolerance. Equations~\eqref{eq:eta}--\eqref{eq:uniform_criterion} make that prescription quantitative: the omitted-space residual provides a conservative sufficient error test for a specified Hamiltonian, while Eq.~\eqref{eq:nested_criterion} supplies a less conservative but noncertifying basis-comparison diagnostic. The latter is what our NaCs calculations presently support. Our results establish conditional stability for specified selected-space coherent calculations, not a universal accuracy standard. For the matched complete-channel candidates, the nonnegative decomposition in Eq.~\eqref{eq:leakage_decomposition} further shows that final logical leakage accounts for most of the nominal infidelity; it does not identify an intrinsic performance limit. In particular, they do not determine the error against the actual experiment because the microwave polarization, some line-power conditions, molecular motion, pulse-transfer function, and loss channels are not independently constrained. The experimental truth-table and Bell-state figures of merit reported by Picard \emph{et al.}~\cite{Picard2025} should not be directly compared with Eq.~\eqref{eq:metric}.

These observations motivate the broader question of \emph{identifiability}: which spectroscopic and dynamical observations determine the effective Hamiltonian components relevant to the desired measurement? We have not constructed distinct Hamiltonians that fit the same observations within their uncertainties while predicting different gates, and therefore do not establish Hamiltonian non-identifiability. What the calculations directly show is that target-transition calibration and complete-gate calibration are distinct tasks. Our calculations suggest a discriminating experimental strategy, but do not yet constitute its validated prediction. In RbCs pairs, Ruttley \emph{et al.} fitted single-molecule and pair spectra together with Rabi oscillations using shared parameters. The pair data resolved an interaction shift and a collective Rabi rate enhanced by $\sqrt{2}$ relative to the single-molecule rate~\cite{Ruttley2025}. For the NaCs storage gate, the corresponding open question is which measurements distinguish spectator excitation from a target-frequency shift. One could scan tweezer ellipticity while repeatedly calibrating the intended target line and Rabi rate, then measure the exchange dynamics, stage-specific nonlogical populations, and a phase-sensitive logical observable. Measuring the exchange rate separately would help distinguish an interaction-timing mismatch from excitation into spectators; the usefulness of that distinction remains to be tested quantitatively. A reduced model that tracks only its intended transition and an expanded multilevel model need not predict the same residual response. Quantifying the difference with uncertainty bands on microwave polarization would be an especially valuable continuation, requiring new experimental input and separate calculations.

\appendix
\section{Derivation and scope of the residual bound}
\label{app:proof}
This appendix records the short proof of Eqs.~\eqref{eq:state_error_bound}--\eqref{eq:fidelity_bound} so that the assumptions are explicit. Set $V(t)=U_\Pi(t,0)L$, embedded in the complete Hilbert space. By construction, $V(0)=L$ and $i\hbar\dot V=\Pi H\Pi V$. Hence its residual with respect to the full Schr\"odinger equation is
\begin{equation}
(i\hbar\partial_t-H)V(t)=-QH(t)\Pi V(t)=-R_L(t).
\end{equation}
Variation of constants (Duhamel's formula) gives
\begin{equation}
U(T,0)L-V(T)=-\frac{i}{\hbar}\int_0^T U(T,t)R_L(t)\,dt.
\label{eq:duhamel}
\end{equation}
Since $U(T,t)$ is unitary, its left multiplication does not increase
the operator norm. Taking the triangle inequality proves
Eq.~\eqref{eq:state_error_bound}. Multiplication by $L^\dagger$ and
$\|X\|_F\le\sqrt d\,\|X\|_2$ proves Eq.~\eqref{eq:map_bound}.

For completeness, let $M$ and $N$ be two $d\times d$ contractions and
$W$ a unitary target, with a fixed permitted output correction $Z$.
The expression in Eq.~\eqref{eq:metric} obeys
\begin{equation}
\left|F_{\rm avg}(M;Z)-F_{\rm avg}(N;Z)\right|
\le\frac{2}{\sqrt d}\,\|M-N\|_F,
\end{equation}
using $\|M\|_F,\|N\|_F\le\sqrt d$,
$|\Tr(W^\dagger ZM)|\le d$, and
$|\Tr[W^\dagger Z(M-N)]|\le\sqrt d\,\|M-N\|_F$.
Maximizing over the same allowed set of $Z$ on each side retains the
inequality. The proof requires a fixed projector $\Pi$, common
initial-state embedding, Hermitian full dynamics, and matched waveform
and frame conventions. It does not address numerical time-discretization
error, uncertain Hamiltonian parameters or a time-dependent projector;
the latter would introduce geometric couplings involving $\dot\Pi$.

\section{Effective optical fit and seven-line accounting}
\label{app:fit}
The one-molecule calculation uses the $N=0,1$ internal basis and the theoretical NaCs constants of Ref.~\cite{Aldegunde2017}; these constants and $B=\SI{864.1}{\gauss}$ were held fixed. The effective optical tensor contribution is parameterized by $u_t$ and the polarization vector $\bm\epsilon_{\rm opt}=(\cos\chi,i\sin\chi,0)$. Only these two parameters are fitted. The angle $\chi=54.817^\circ$ is a coordinate convention for the complex polarization components, not a directly measured conventional ellipse angle or a laboratory-axis reconstruction. A common scalar Stark shift is immaterial to the transition offsets considered here.

Let $\mathcal B=\{0\to e,0\to f_z,0\to f_y\}$ and let $\nu_i^{\rm calc}$ be the static eigenenergy difference expressed as an offset from $2B_{\rm rot}$. The implemented least-squares objective is
\begin{equation}
S(u_t,\chi)=\sum_{i\in\mathcal B}\left[\nu_i^{\rm calc}(u_t,\chi)-\nu_i^{\rm obs}\right]^2.
\label{eq:optical_fit_cost}
\end{equation}
This is an \emph{unweighted} sum of squared residuals in kHz$^2$, rather than an uncertainty-normalized $\chi^2$ function; the quoted line-center standard errors therefore do not produce parameter confidence intervals. The implementation uses \texttt{scipy.optimize.least\_squares}, initialized at $u_t/h=-3500$ kHz and $\chi=54.7^\circ$, with bounds $-6000\le u_t/h\le-1000$ kHz and $48^\circ\le\chi\le85^\circ$. At each trial the 128-dimensional Hamiltonian is diagonalized and the three bright-line offsets are recomputed. The fitted effective parameters are $u_t/h=-3526.155$ kHz and $\chi=54.817^\circ$ at the reference field.

The absolute calculated offsets of these three lines have nearly equal positive residuals, about $5.88$ kHz, relative to the published $2B_{\rm rot}$ reference. The empirical $-5.88$ kHz common-reference correction is applied \emph{only to the reporting columns} in Table~\ref{tab:sevenlines}: it was not fitted as a third Hamiltonian parameter or included in the gate energies. Consequently, the apparent agreement of the \emph{relative} bright-state splittings is an in-sample result, not a proof that the absolute rotational reference is reproduced. The held-out $0\to f_w$, $g_+\to e$, $g_-\to e$ and $1\to e$ lines remain discrepant in the static spectrum after this common correction.

The optical fit is implemented in \texttt{nacs.spectroscopy.calibrate}. Table~\ref{tab:sevenlines} uses static transition frequencies calculated from the same dressed molecular Hamiltonian as the gate model.

A separate finite-power microwave calculation can shift apparent resonance maxima, but its adjustable polarization and line-specific powers were inferred from observed discrepancies. Those inferred shifts are neither included in Table~\ref{tab:sevenlines} nor counted as out-of-sample validation. Without a measured set of line centers versus calibrated microwave power and polarization, the static model's residuals cannot be interpreted uniquely as either incorrect molecular constants or finite-power dressing.

\section{Numerical comparisons}
The sensitivity analysis compares three microwave calibration conventions for each perturbed Hamiltonian. The basis analysis ranks candidates from all 128 molecular eigenstates and propagates up to 25 states per molecule. Lines connecting calculated points in the figures guide the eye.

\section{What the calculations do not establish}
We do not propagate the entire $16{,}384$-dimensional pair basis within the chosen $N=0,1$ manifold. Even propagation in that space would retain the rotational truncation, secular approximation, and fixed molecular positions. The screening criteria do not bound all multistep virtual transitions. Neither optical field parameters nor microwave polarization have a complete experimental calibration, and hypothesized finite-power shifts are not independent confirmation of the spectral fit. We do not include quantized trapping motion, dissipation, detection errors, or realistic apparatus bandwidth measurements. We do not demonstrate superiority to the robust effective-model control theory of Ref.~\cite{Bergonzoni2025} or the alternative gate of Ref.~\cite{LuShi2026}. The available coherent fidelity numbers are conditional model benchmarks and should not be equated with experimental entanglement or truth-table fidelities.

\end{document}